\documentclass[twocolumn]{jpsj2} 

\title{Genuine Phase Diagram of Homogeneously Doped CuO$_2$ Plane\\ in High-$T_c$ Cuprate Superconductors}

\author{Hidekazu Mukuda$^{1}$\thanks{E-mail address: mukuda@mp.es.osaka-u.ac.jp}, Yuhei Yamaguchi$^{1}$, Sunao Shimizu$^{1}$, Yoshio Kitaoka$^{1}$, Parasharam Shirage$^{2}$, Akira Iyo$^{2}$ }

\inst{$^{1}$Graduate School of Engineering Science, Osaka University, Toyonaka, Osaka 560-8531 \\
$^2$National Institute of Advanced Industrial Science and Technology (AIST), Umezono, Tsukuba 305-8568 \\
}

\abst{
We report a genuine phase diagram for a disorder-free CuO$_2$ plane based on the precise evaluation of the local hole density ($N_h$) by site-selective Cu-NMR studies on five-layered high-$T_c$ cuprates. It has been unraveled that (1) the antiferromagnetic metallic state (AFMM) is robust up to $N_h\approx$ 0.17, (2) the uniformly mixed phase of superconductivity (SC) and AFMM is realized at $N_h\le$ 0.17, (3) the tetracritical point for the AFMM/(AFMM+SC)/SC/PM(Paramagnetism) phases may be present at $N_h \approx$ 0.15 and $T \approx$ 75 K, (4) $T_{\rm c}$ is maximum close to a quantum critical point (QCP) at which the AFM order collapses, suggesting the intimate relationship between the high-$T_{\rm c}$ SC and the AFM order. The results presented here strongly suggest that the AFM interaction plays the vital role as the glue for the Cooper pairs, which will lead us to a genuine understanding of why the $T_{\rm c}$ of cuprate superconductors is so high. 
}

\kword{superconductivity, copper-oxide, antiferromagnetism, NMR, phase diagram}

\begin{document}
\maketitle

\date{\today}

\section{Introduction}

\begin{figure}[htbp]
\begin{center}
\includegraphics[width=1\linewidth]{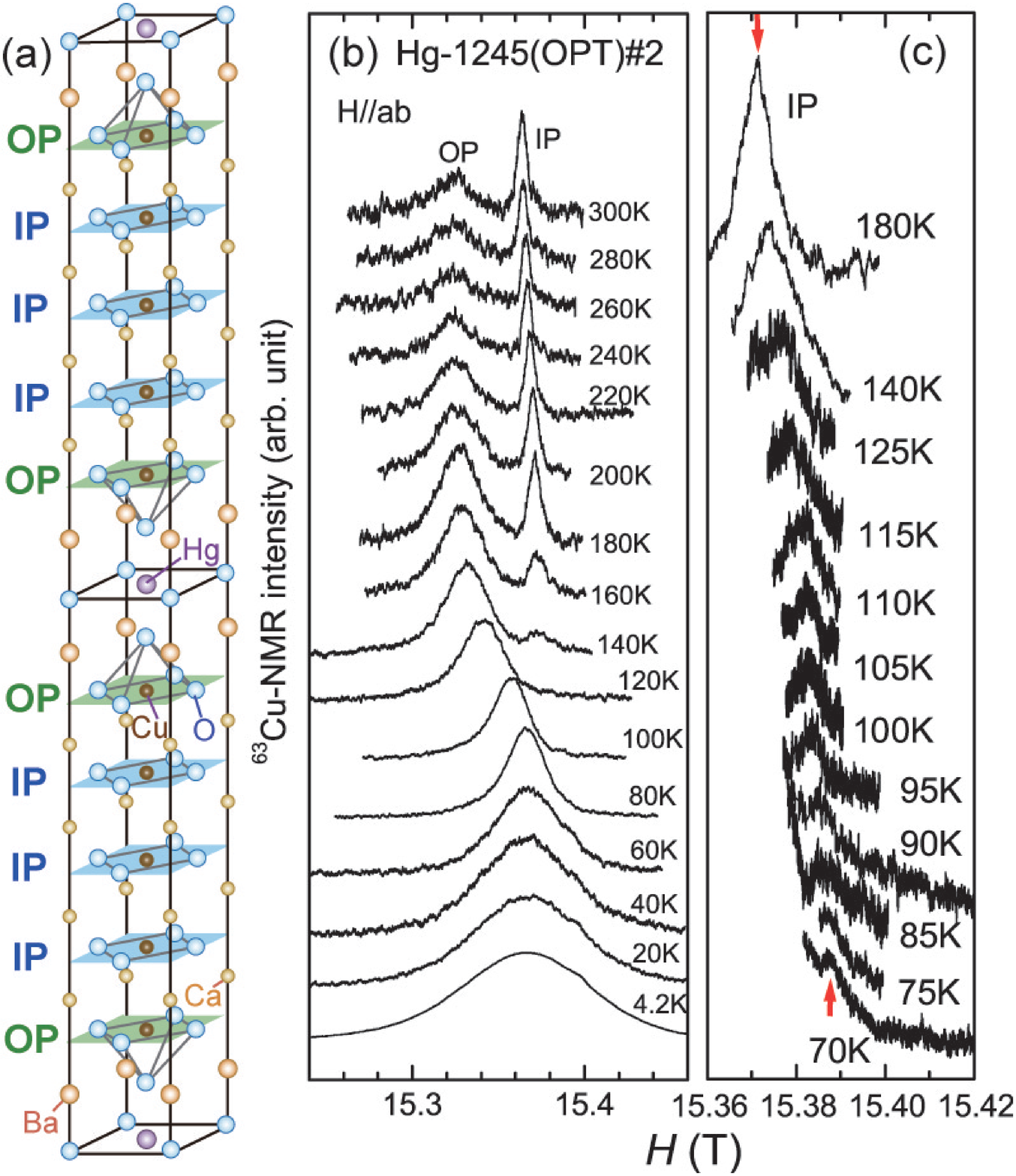}
\end{center}
\caption{(Color online) (a) Crystal structure of five-layered Hg-1245. (b)Temperature dependence of $^{63}$Cu-NMR spectra of Hg-1245(OPT)$\sharp2$. The $^{63}$Cu-NMR signals from a pyramid-type outer CuO$_2$ plane (OP) and a square-type inner plane (IP) are separately observed due to the difference of the Knight shift and nuclear quadrupole shift. The narrow linewidths in the Cu NMR spectra, particularly for IPs, indicate disorder-free CuO$_2$ planes that are homogeneously doped. 
It should be noted that the resonance peak of IP disappears below $\sim$130K due to the short nuclear spin-spin relaxation time $T_2$ at IP site. (c) However, in the appropriate pulse condition, we can detect the extremely small resonance signal of IP down to 70K, which enables us to determine Knight shift for IP between 70 K and 300 K. }
\label{fig:spectra}
\end{figure}

Despite of more than 22 years of research, there is still no universally accepted theory for mechanism of cuprate superconductors. 
The high-$T_c$ superconductivity (SC) in cuprates emerges on a CuO$_2$ plane when an antiferromagnetic (AFM) Mott insulator is doped with mobile carriers. 
A strong relationship between AFM order and SC is believed to be a key to understand the origin of their remarkably high SC transition \cite{Anderson,Inaba,Ogata,Demler,Moriya}. 
Experimentally, however, in a prototype high-$T_c$ cuprate La$_{2-x}$Sr$_x$CuO$_4$ (LSCO), the AFM and SC phases are separated by the spin-glass phase in association with the carrier localization \cite{Keimer}. 
Since chemical substitution is necessary for doping, a disorder effect poses a significant and inevitable problem in doped cuprates generally. 
Multilayered cuprates provide us with the opportunity to research the characteristics of the disorder-free CuO$_2$ plane. 
Figure \ref{fig:spectra}(a) shows the crystal structure of the Hg-based five-layered cuprate HgBa$_2$Ca$_4$Cu$_5$O$_{12+\delta}$(Hg-1245) composed of two types of CuO$_2$ planes in a unit cell: a pyramid-type outer CuO$_2$ plane (OP) and a square-type inner plane (IP) \cite{IPs}. 
The site-selective NMR is the best and the only tool used to extract layer-dependent characteristics \cite{Tokunaga,Kotegawa2001,Kotegawa2004,MukudaPRL2006,MukudaJPSJ2006}. 
Since the IPs are farther from the charge reservoir layers (HgO$_\delta$) than the OPs, the carrier density at IPs is lower than that at OPs. 
The disorder introduced along with the chemical substitution in an HgO$_\delta$ layer is effectively shielded on an OP, as a result of which ideally flat CuO$_2$ planes are realized, especially at IPs, differentiating multilayered cuprates from mono-layered cuprate LSCO. 
In particular, the uniform mixing state of SC and AFM order on an OP for underdoped Hg-1245 has been shown in a previous study \cite{MukudaPRL2006}; however, the carrier density and N\'eel temperature inherent in this layer have not been identified. 

In this paper, we have revealed the intrinsic phase diagram of a disorder-free CuO$_2$ plane on the basis of the site-selective Cu-NMR studies on Hg-based five-layered cuprates. 
The most remarkable feature of this phase diagram is that the AFM metallic phase is robust and coexists with the SC phase in a region extending up to the optimally doped region. $T_c$ has a peak close to a quantum critical point at which the AFM order collapses, suggesting the intimate relationship between AFM order and SC. Our findings imply that the AFM interaction plays the vital role as the glue for the Cooper pairs. 

\section{Experimental}

A polycrystalline sample of Hg-1245(OPT)$\sharp2$ has been synthesized at a pressure (temperature) of 2.5 GPa (950$^\circ$C), which is lower than that applied in the sample synthesis in the previous studies \cite{Tokiwa_sample,Kotegawa2004}, i.e., 4.5 GPa (1050$^\circ$C), and referred to as "Hg-1245(OPT)$\sharp1$" in this study. 
The SC transition temperature $T_{\rm c}$ has been determined to be 110 K from the onset of a sharp diamagnetic signal in dc susceptibility. 
For the NMR measurements, we used oriented powder samples aligned along the c-axis under a high magnetic field. 
The narrow linewidths in the Cu NMR spectra, particularly those less than 50 Oe for IPs even at 15 T($H \| c$), indicate disorder-free CuO$_2$ planes that are homogeneously doped, which enables us to precisely measure the Knight shift. 

\section{Results}

\subsection{Superconducting property of Hg-1245(OPT)$\sharp2$}

Figure \ref{fig:spectra}(b) shows temperature dependence of field swept $^{63}$Cu NMR spectra obtained at a fixed frequency of 174.2MHz in the field perpendicular to the c axis. 
The $^{63}$Cu-NMR signals from OP and IP are separately observed due to the difference of the Knight shift and nuclear quadrupole shift \cite{Kotegawa2004}. 
The narrow linewidths in the Cu NMR spectra, particularly for IPs, indicate disorder-free CuO$_2$ planes that are homogeneously doped, and also imply that the disorder introduced along with the chemical substitution in an HgO$_\delta$ layer is effectively shielded on an OP. 
It should be noted that the resonance peak of IP disappears below $\sim$130K, as shown in Fig.\ref{fig:spectra}(b), due to the short nuclear spin-spin relaxation time $T_2$ at IP site. 
However, in the appropriate pulse condition that the interval of the $\pi/2$ and $\pi$ pulse is as short as 3.4$\mu$sec, we succeeded in detecting the extremely small resonance signal of IP down to 70K, as shown in Fig. \ref{fig:spectra}(c), which enables us to determine Knight shift for IP between 70 K and 300 K. 

\begin{figure}[htbp]
\begin{center}
\includegraphics[width=1\linewidth]{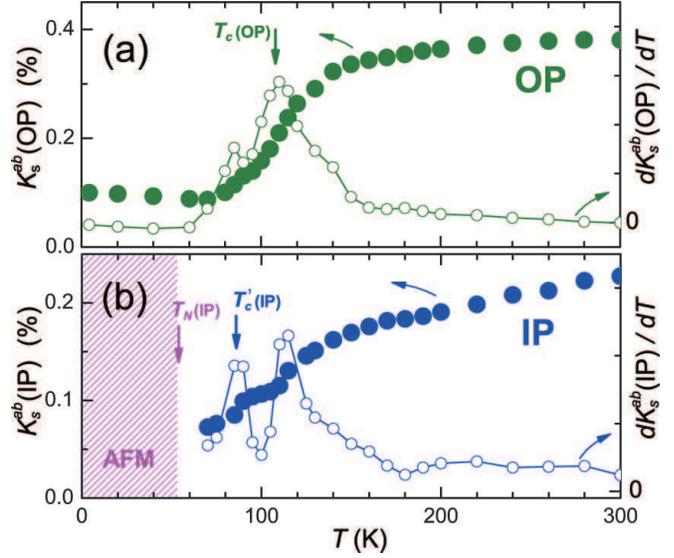}
\end{center}
\caption{(Color online) Temperature dependence of $K_s^{ab}$ for (a) OP and (b) IP of Hg-1245(OPT)$\sharp2$. The solid and empty circles represent $K_s$ and its temperature-derivatives, respectively. $K_s^{ab}$(OP) decreases rapidly below $T_{\rm c}$ = 110 K, whereas $K_s^{ab}$(IP) decreases significantly at $T =$ 85 K. This indicates that although the bulk SC transition is driven primarily by OPs, the SC gap in IPs rapidly develops below $T_{\rm c}$'(IP)$=$ 85 K, that is inherent $T_{\rm c}$ of IPs. }
\label{fig:K_Hg}
\end{figure}

\begin{figure}[htbp]
\begin{center}
\includegraphics[width=1\linewidth]{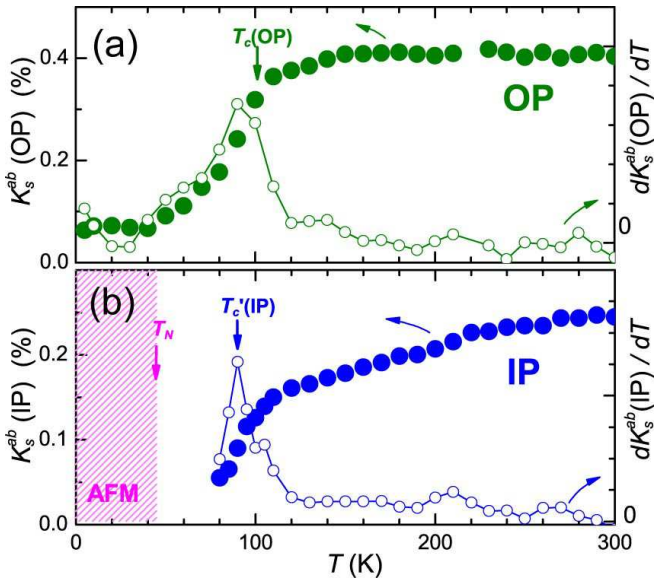}
\end{center}
\caption{(Color online) Knight shift ($K_s^{ab}$) measurement for (a) OP and (b) IP of slightly overdoped Tl-1245(OVD) with $T_{\rm c}$ = 100 K. The solid and empty circles represent $K_s$ and its temperature derivatives, respectively. In this study, we have detected small resonance signal of IP down to 80 K in the appropriate pulse condition. It has been found that the SC transition inherent in IPs takes place at $T_{\rm c}$'(IP)$=$ 90($\pm$5) K, and the bulk SC transition at 100 K is driven primarily by OPs. The AFM order emerges at IPs below $T_N = 45$ K\cite{Kotegawa2004} with a moment of $M_{\rm AFM}$(IP)$\sim$0.1$\mu_{\rm B}$, whereas the OP is nearly paramagnetic\cite{MukudaPRL2006}. }
\label{fig:K_Tl}
\end{figure}

The Knight shift generally consists of spin and orbital components denoted as $K_s$ and $K_{orb}$, respectively. 
Figure \ref{fig:K_Hg} shows the temperature ($T$) dependences of $K_s^{ab}$ ($K_s^{ab}$ in a field parallel to the ab-plane) for OPs and IPs of Hg-1245(OPT)$\sharp2$, respectively. Here, $K_s^{ab}$ is obtained by subtracting temperature-independent $K_{orb}^{ab}$, which is approximately 0.2-0.21\%, irrespective of IP or OP for Hg-based cuprates \cite{Julien,Magishi}. 
$K_s^{ab}$ of OPs decreases rapidly below $T_{\rm c}=$ 110 K. 
We note that a distinct peak in the temperature derivatives of $K_s^{ab}$(OP) coincides with $T_{\rm c}=$ 110 K. 
On the other hand, $K_s^{ab}$(IP) decreases significantly at $T <$ 85 K in addition to its observed decrease at bulk $T_{\rm c}$ of 110 K, which has been corroborated by the two peaks at 85 and 110 K in the temperature derivatives of $K_s^{ab}$(IP). 
This result reveals that the bulk SC transition is driven primarily by an optimally doped OP, but the SC transition inherent in underdoped IPs manifests itself at $T_{\rm c}$'(IP) $\approx$ 85($\pm$5) K due to a large imbalance in the carrier densities between OPs and IPs. 
It is naturally expected that IPs exhibit superconductivity between 85 and 110 K due to the proximity effect, which has been observed in other multilayered cuprates \cite{Tokunaga,Kotegawa2001}. 
The analogous behavior was also observed in slightly overdoped Tl-1245(OVD) with $T_N = 45$ K \cite{Kotegawa2004}, in which the SC transition inherent in IPs is $T_{\rm c}$'(IP) $\approx$ 90($\pm$5) K, whereas the bulk SC transition at 100 K is driven primarily by OPs, as shown in Fig. \ref{fig:K_Tl}.
As discussed in section 3.3, it is reasonable that $T_N$ is lower and $T_c$ is higher  than those of Hg-1245(OPT)$\sharp2$, since the carrier density of IPs is slightly higher than that for Hg-1245(OPT)$\sharp2$.

\begin{figure}[htbp]
\begin{center}
\includegraphics[width=1\linewidth]{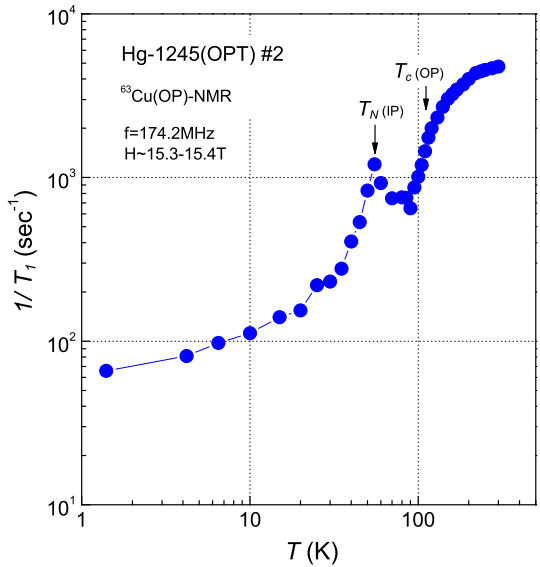}
\end{center}
\caption{(Color online) Nuclear-spin relaxation rate $1/T_1$ for the OPs of Hg-1245(OPT)$\sharp2$.  The N\'eel temperature $T_{\rm N}$ = 55 K at IPs has been identified by a distinct peak due to the critical slowing down in the plot of $1/T_1$ versus $T$ at OP. Similarly, in a previous paper\cite{Kotegawa2004}, $T_{\rm N}$ for Hg-1245(OPT)$\sharp1$ and Tl-1245(OVD) has been identified to be 60 and 45 K, respectively. Here, as for in the SC state, the short component of $T_1$ is plotted in the figure since it is sensitive to the magnetic order at IPs \cite{Kotegawa2004}. }
\label{fig:T1}
\end{figure}

\subsection{Magnetic property of Hg-1245(OPT)$\sharp2$}

\begin{figure}[htbp]
\begin{center}
\includegraphics[width=1\linewidth]{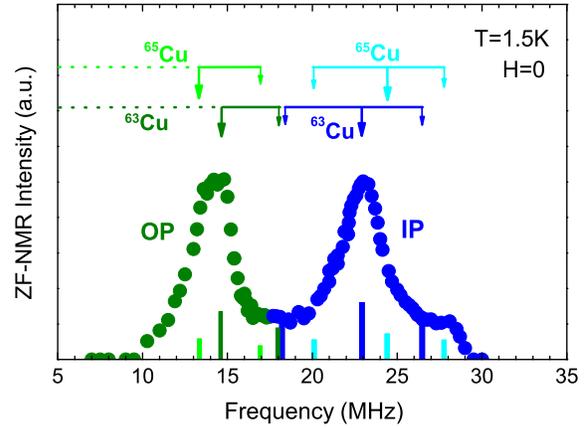}
\end{center}
\caption{(Color online)  Zero-field Cu NMR spectrum of Hg-1245(OPT)$\sharp2$ at 1.5 K gives a strong evidence of the onset of the AFM order at IPs. The spectrum around 23 MHz is reproduced by assuming the internal field to be 2.0 T, which is induced by the spontaneous moments ($M_{\rm AFM}$(IP)$\sim$0.1$\mu_{\rm B}$) produced by the AFM order at the Cu sites in IPs. The bars indicate the calculated resonance lines in IP and OP for $^{63,65}$Cu isotopes. The peak at 15 MHz has been attributed to OPs in the paramagnetic state because the peak almost coincides with the NQR frequency for OPs, $^{63}\nu_Q$(OP)$=$ 16 MHz \cite{Kotegawa2004}. The slight shift to lower frequency side derives from a presence of a tiny field ($\le 0.1$ T) at the OP due to the proximity effect from IPs. }
\label{fig:ZFNMR}
\end{figure}

As presented in Fig. \ref{fig:spectra}(c), the NMR signal of IPs of Hg-1245(OPT)$\sharp2$ disappears below 70 K because of the extremely short relaxation time due to the development of critical AFM spin fluctuations towards a possible N\'eel ordering $T_N$ even in the SC state. 
Generally a peak of nuclear-spin relaxation rate $1/T_1$ is observed at $T_N$ due to critical slowing down. 
In fact, as shown in Fig. \ref{fig:T1}, $T_{\rm N}=$ 55 K has been confirmed by a peak in a plot of $1/T_1$ versus $T$ at an OP. 
The onset of the AFM order at IPs has been evidenced by the zero-field (ZF) Cu NMR spectrum at 1.5 K without any external field, as shown in Fig. \ref{fig:ZFNMR}. 
The peak at 15 MHz has been attributed to OPs in the paramagnetic state because the peak almost coincides with the nuclear quadrupole resonance (NQR) frequency for OPs, $^{63}\nu_Q$(OP)$=$ 16 MHz \cite{Kotegawa2004}. The slight shift to lower frequency side may derive from a presence of a tiny field ($\le 0.1$ T) at the OP due to the proximity effect from IPs.
The spectrum of IPs is observed at 23 MHz, not at $^{63}\nu_Q$(IP)$=$ 8.4 MHz \cite{Kotegawa2004}. 
The spectral analysis of IPs, assuming a Zeeman field, reveals that an internal field ($H_{\rm int}$) of 2.0 T is induced by the spontaneous AFM moments $M_{\rm AFM}$(IP) due to the AFM order at the Cu site in IPs, as displayed by the bars in the figure. 
The unique value of $M_{\rm AFM}$(IP) is evaluated to be 0.095$\mu_{\rm B}$ per Cu site at IPs by using the relation $H_{\rm int}$(IP)$= |A_{\rm hf}$(IP)$|M_{\rm AFM}$(IP) with the hyperfine coupling constant $A_{\rm hf}$(IP)$=-$ 20.7 T$/\mu_{\rm B}$ \cite{Kotegawa2004}. 
It is remarkable that this AFM moment spontaneously emerges at superconducting IPs with a possible commensurate AFM structure. Here, we can exclude the spin-glass state at IPs because the internal field at IPs ($H_{\rm int}$ = 2.0 T) is almost the same at all the Cu(IP) sites and its distribution is less than $\pm$ 0.2 T. These results provide the microscopic evidence of the uniform mixing of SC and AFM order on disorder-free IPs with an indisputable moment. 

\begin{figure}[htbp]
\begin{center}
\includegraphics[width=1\linewidth]{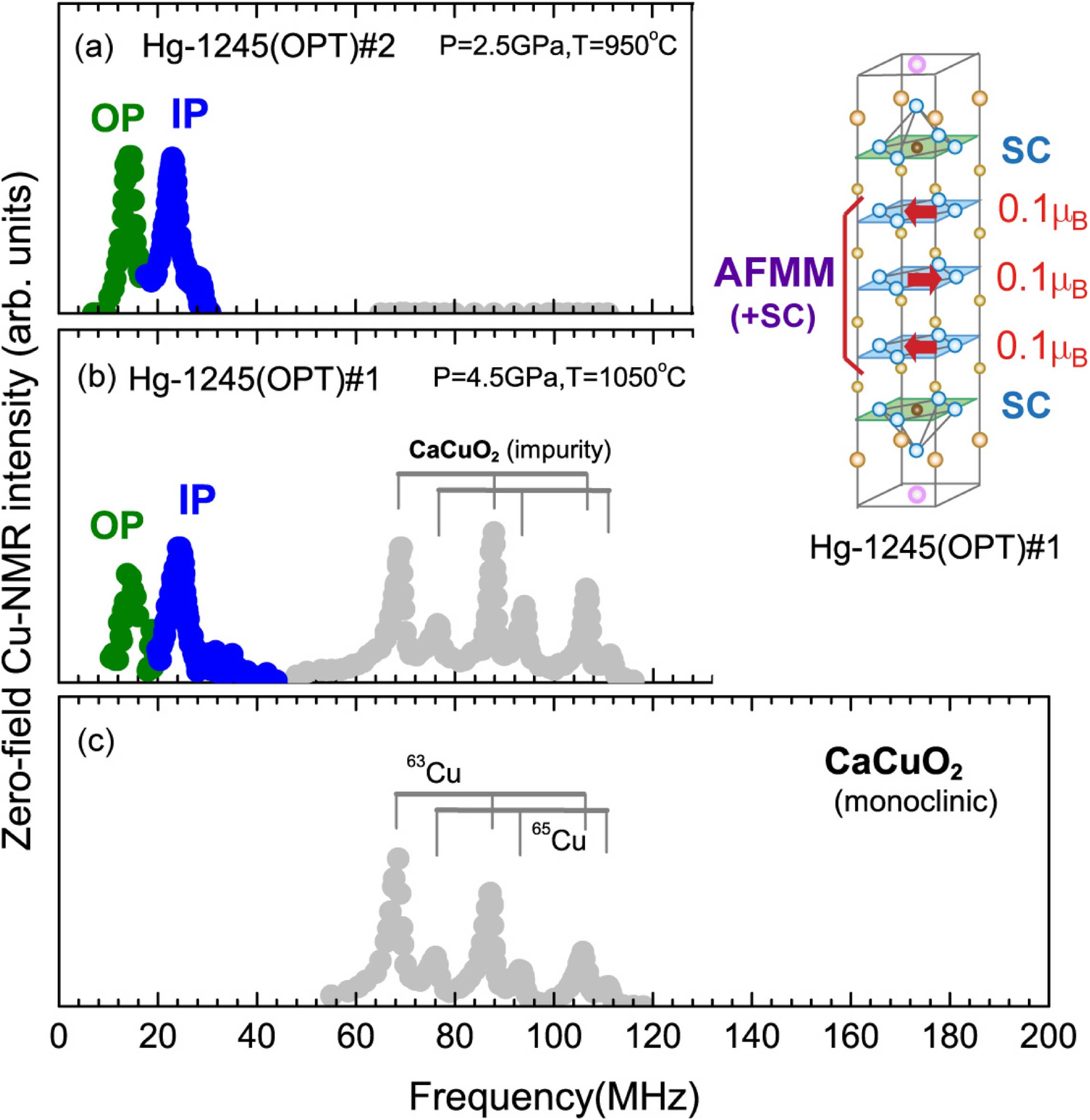}
\end{center}
\caption{ (Color online) (b)In the previous study on the Hg-1245(OPT)$\sharp1$ \cite{Kotegawa2004}, the zero-field NMR spectrum between 50-120 MHz were assigned to be IPs, which has AFM moments of 0.3-0.37$\mu_{\rm B}$.  
However, this spectrum has been identified to be the spectrum of the impurity phase, CaCuO$_2$(II) with a monoclinic structure \cite{Takano} by comparing the spectrum with that of (c) pure CaCuO$_2$(II) sample. This impurity phase is produced only during the synthesis at a high pressure (high temperature) of 4.5 GPa (1050$^\circ$C)\cite{Takano}; hence (a) it has not been observed in the present study on Hg-1245(OPT)$\sharp2$ synthesized under a pressure (temperature) of 2.5 GPa (950$^\circ$C). As a result, the peak at 24 MHz for Hg-1245(OPT)$\sharp1$ is assigned to be IPs with the moment of $M_{\rm AFM}$(IP)$\approx0.1\mu_{\rm B}$. This value is comparable to that observed in Hg-1245(OPT)$\sharp2$, which is consistent with that the carrier density of IPs is similar each other. }
\label{fig:CaCuO2}
\end{figure}

In the previous study on the Hg-1245(OPT)$\sharp1$ \cite{Kotegawa2004},  the AFM moment at IP site of this compound were evaluated to be 0.3-0.37$\mu_{\rm B}$ from the spectrum analysis between 50-120 MHz(Fig. \ref{fig:CaCuO2}(b)). However, through the systematic NMR  investigations on the various multilayered cuprates in the several years, the spectrum between 50-120 MHz has been identified to be the spectrum of the impurity phase, CaCuO$_2$(II) with a monoclinic structure \cite{Takano} by comparing the spectrum with that of pure CaCuO$_2$(II) sample, as shown in Fig. \ref{fig:CaCuO2}(c). 
This impurity phase is produced only during the synthesis at a high pressure (high temperature) of 4.5 GPa (1050$^\circ$C)\cite{Takano}; hence it has not been observed in the present study on Hg-1245(OPT)$\sharp2$ synthesized under a pressure (temperature) of 2.5 GPa (950$^\circ$C), as indicated in Fig. \ref{fig:CaCuO2}(a). 
As a result, the peak at 24 MHz for Hg-1245(OPT)$\sharp1$ is assigned to be IPs with the moment of $M_{\rm AFM}$(IP)$\approx0.1\mu_{\rm B}$, and the other peak at $\sim$14 MHz is assigned to be OP in the paramagnetic state that is affected by a very tiny field due to the proximity effect from IPs.
Thus, we note that the magnitude of $M_{\rm AFM}$(IP)$\approx 0.1\mu_{\rm B}$ of Hg-1245(OPT)$\sharp2$ is comparable to that not only in Tl-1245(OVD) \cite{MukudaPRL2006} but also in Hg-1245(OPT)$\sharp1$, which is consistent with that the carrier density of IPs is similar each other, as discussed in next section.

\subsection{Evaluation of the carrier density}

\begin{figure}[htbp]
\begin{center}
\includegraphics[width=1\linewidth]{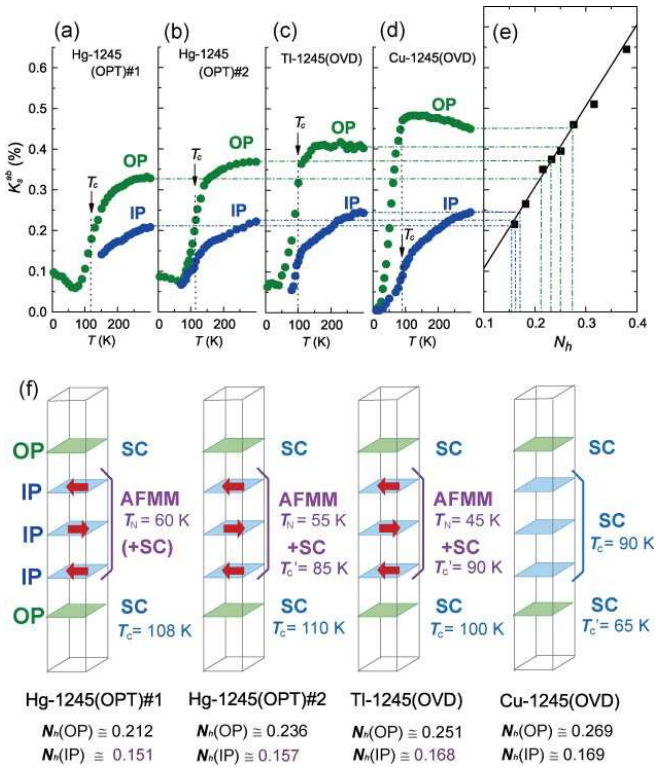}
\end{center}
\caption{(Color online) Temperature dependence of $K_s^{ab}$ for (a) Hg-1245(OPT)$\sharp1$[cited from Ref.\cite{Kotegawa2004}], (b) Hg-1245(OPT)$\sharp2$, (c) Tl-1245(OVD), and (d) Cu-1245(OVD)[cited from Ref.\cite{Kotegawa2001}]. (e) $K_s^{ab}$(RT) at room temperature is proportional to the local carrier density $N_h$ [cited from Refs.\cite{Zheng,Kotegawa2001}], which enables us to estimate $N_h$ for IPs and OPs independently. We have shown that the AFM order spontaneously emerges at disorder-free IPs with $N_h$ = 0.15-0.168 below $T_N=$ 60-45 K, whereas the SC transition temperature inherent in these layers is $T_c($IP$) \approx$ 85-90 K. In contrast, the AFM order is not observed in IPs with $N_h\approx$ 0.169 of (d) Cu-1245(OVD), suggesting that a quantum critical point (QCP) exists at $N_h\approx$ 0.17 in the five-layered cuprates. The typical error in $N_h$ is $\pm$0.005.
(f) Layer-dependent physical properties for various five-layered cuprates are summarized. 
 Here, the value of $M_{\rm AFM}$(IP) is almost 0.1$\mu_{\rm B}$ for Hg-1245(OPT)$\sharp1$, Hg-1245(OPT)$\sharp2$ and Tl-1245(OVD). 
}
\label{fig:Carrier}
\end{figure}

In the previous papers\cite{Kotegawa2004,MukudaPRL2006}, the carrier estimation of the underdoped region had remained as a problem, since the $N_h$(IP) for Hg-1245(OPT)$\sharp1$ was tentatively inferred from the whole carrier density evaluated by Hall coefficient and the $N_h$(OP) by Knight shift.
In this study, we evaluate the local carrier densities $N_h$ for these layers only from the Knight shift on the basis of the systematic NMR measurements on the various five-layered cuprates. 
As shown in Fig. \ref{fig:Carrier}(e), it has been established that $N_h$ in various cuprates can be experimentally deduced from the value of $K_s^{ab}$ at room temperature by using the linear relation $N_h$ = 0.0462 + 0.502$K_s^{ab}$(RT)\cite{Zheng,Kotegawa2001}. 
Note that this relation is valid for various high-$T_{\rm c}$ cuprates, irrespective of the type (square or pyramid) and/or number of CuO$_2$ planes. In fact, the carrier densities at OPs and IPs have been independently evaluated to be $N_h$(OP)=0.236 and $N_h$(IP)=0.157 for Hg-1245(OPT)$\sharp2$. We have summarized the physical properties and $N_h$ estimated from the Knight shift of various five-layered cuprates and shown in Figs. \ref{fig:Carrier}(a-f). 
It has been demonstrated that the AFM ordering takes place at IPs at least in the range of $0.151 \le N_h  \le 0.168$. 
The values of $M_{\rm AFM}$(IP)$\sim0.1\mu_{\rm B}$ are observed for these cuprates, which are significantly reduced by the mobile holes from 0.5-0.7$\mu_{\rm B}$ in undoped cuprates \cite{MukudaPRL2006,Vaknin}. 
This emphasizes that the AFM metallic (AFMM) phase persists at IPs, although the doping level is as high as $N_h <$ 0.17. In contrast, the AFM ordering is not observed at IPs with $N_h$ = 0.169 in Cu-1245(OVD)\cite{Kotegawa2001}, suggesting that a quantum critical point (QCP) exists at $N_h\approx$ 0.17 in the five-layered cuprates. 

\begin{figure}[htbp]
\begin{center}
\includegraphics[width=1\linewidth]{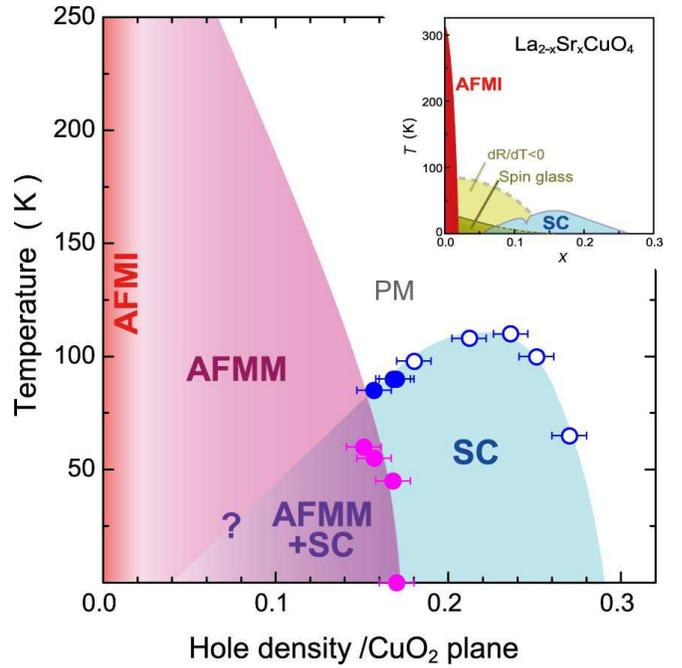}
\end{center}
\caption{(Color online) Phase diagram of homogeneously doped CuO$_2$ plane. On the basis of the results of various five-layered cuprates, $T_N$ and $T_c$ are plotted as functions of $N_h$ per CuO$_2$ plane determined only from the Knight shift. The solid and empty circles correspond to the data for IPs and OPs, respectively. It has been shown that (1) the AFM metallic (AFMM) phase is robust up to $N_h \sim$0.17, (2) the uniformly mixed state of SC and AFMM is realized at least in 0.14 $< N_h \le$ 0.17, (3) $T_c$ has a peak close to the QCP at which the AFM order collapses, indicating the strong relationship between the high-$T_c$ SC and AFM order. This phase diagram differs significantly from the well-established phase diagram of LSCO [cited from Ref.\cite{Keimer}] (see inset), in which both the phases are separated by the spin-glass phase in association with the carrier localization given by $d(resistivity)/dT <0$. 
}
\label{fig:PhaseDiagram}
\end{figure}

\section{Discussions}

\subsection{Phase diagram established in five-layered cuprates}

We obtained the novel phase diagram by plotting $T_{\rm N}$ and $T_{\rm c}$ as functions of $N_h$ evaluated only by Knight shift measurement, as presented in Fig. \ref{fig:PhaseDiagram}. 
The characteristic features are summarized as; (1) the AFM metallic state is robust up to $N_h\approx$ 0.17, (2) the uniformly mixed phase of SC and AFMM is realized at 0.15 $< N_h\le$ 0.17 at least, (3) the $T_{\rm c}$ is maximum close to a QCP at which the AFM order collapses, suggesting the intimate relationship between the high-$T_{\rm c}$ SC and the AFM order. 
It is noteworthy that the phase diagram for 0.14 $< N_h <$ 0.18 including the QCP is precisely determined only by homogeneously doped IPs.  This result also indicates the presence of a tetracritical point for the AFMM/[AFMM+SC]/SC/PM(Paramagnetism) phases at $T \approx$ 75 K with $N_h  \approx$0.15 at zero fields. This is the first observation in the phase diagrams of high-$T_{\rm c}$ superconductors, although it was recently reported in the heavy-fermion superconductor CeRhIn$_5$ \cite{YashimaPRB2007}. 
In such cases, it is noted that the SO(5) theory unifies the AFM and SC states by a symmetry principle and describes their rich phenomenology through a single low-energy effective model\cite{Demler} and hence may be applied to describe the quantum phase transition of AFM order and SC involving the tetracritical point. 

\begin{figure}[htbp]
\begin{center}
\includegraphics[width=1\linewidth]{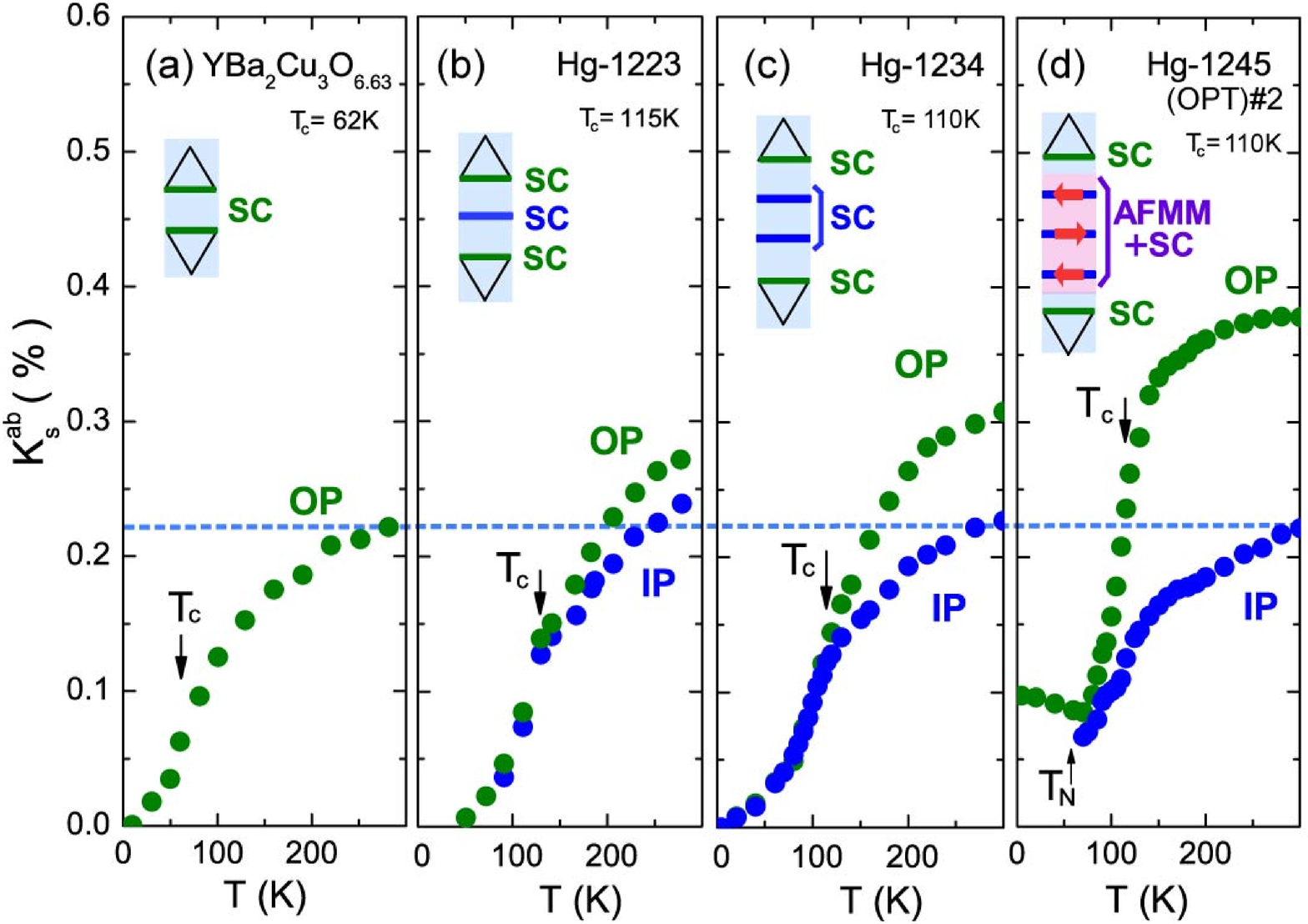}
\end{center}
\caption{(Color online) Knight shift for $n$-layered cuprates; (a) $n = 2$, YBa$_2$Cu$_3$O$_{6.63}$($T_c=$ 62 K) [cited from Ref.\cite{Takigawa}]; (b) $n = 3$, Hg-1223($T_c=$ 115 K) [cited from Ref. \cite{Julien}]; (c) $n = 4$, Hg-1234($T_c=$ 110 K) ; and (d) $n = 5$, Hg-1245(OPT)$\sharp2$. Although their carrier densities are almost $N_h$(OP or IP) = $0.16\pm$0.01 evaluated from $K_s^{ab}$(RT) $\approx$ 0.22\% (dotted line), the ground state still remains paramagnetic for $n$-layered cuprates with $n \le$ 4. This suggests that three underdoped IPs in the case of $n = 5$ may stabilize the long-range AFM order due to sufficient interlayer coupling. We suggest that it is the two-dimensional fluctuations produced by the weak interlayer magnetic coupling that suppresses the AFM order in LSCO and YBCO by doping the small amount of holes. }
\label{fig:Knight_nlayer}
\end{figure}

\begin{figure}[htbp]
\begin{center}
\includegraphics[width=1\linewidth]{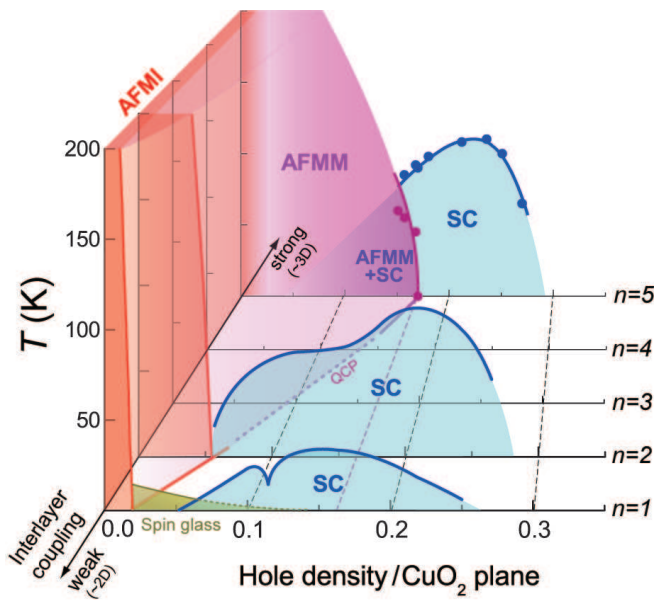}
\end{center}
\caption{(Color online)  Comparison of phase diagrams for $n$-layered cuprates ($n \le 5$). The phase diagram of five-layered cuprates differs from those of LSCO ($n = 1$) and YBCO ($n = 2$), in which the long-range AFM order collapses completely by doping with an extremely small amount of holes. 
In five-layered systems, three underdoped IPs may stabilize the long-range AFM order due to sufficient interlayer coupling, whereas the carrier density at QCP for $n$-layered systems with $n\le 4$ becomes lower than that in five-layered systems because of the weak effective interlayer coupling that may suppress the AFM order. It strongly suggests that the QCP moves to a region of lower carrier density for $n$-layered cuprates with $n \le 4$. Consequently, we consider that the antiferromagnetically coupled spontaneous moment may persist up to $N_h\approx$  0.16 for the CuO$_2$ planes in LSCO and YBCO also; however, it may be hidden within the plane due to the strong 2D fluctuations brought about by the weak interlayer coupling. This concept will lead us to a coherent understanding of underlying anomalies on underdoped cuprates(see text). }
\label{fig:phase_nlayer}
\end{figure}

The phase diagram in Fig. \ref{fig:PhaseDiagram} differs significantly from the well-established phase diagrams of mono-layered LSCO and double-layered YBa$_2$Cu$_3$O$_{6+x}$(YBCO), in which the long-range AFM order collapses completely by doping with an extremely small amount of holes of $N_h\sim$ 0.02 \cite{Keimer} and 0.055 \cite{Sanna}, respectively. 
In fact, we have presented the temperature dependences of $K_s^{ab}$ for underdoped $n$-layered cuprates, e.g., $n=2$: YBa$_2$Cu$_3$O$_{6.63}$ ($T_{\rm c}$ = 62 K) reported by Takigawa {\it et al.}\cite{Takigawa}, $n=3$: Hg-1223 ($T_{\rm c}$ = 115 K) reported by Julien {\it et al.}\cite{Julien}, and $n=4$: Hg-1234 ($T_{\rm c}$ = 110 K), as shown in Fig. \ref{fig:Knight_nlayer}. 
Although their carrier densities are almost the same with $N_h$(OP or IP)$ = 0.16 \pm$ 0.01 evaluated from $K_s^{ab}$(RT)$\approx$0.22\%, it should be noted that the ground state remains still in the paramagnetic state for $n\le 4$. 
Although we have investigated two underdoped Hg-1234 samples ($n=4$) with $T_{\rm c}$ = 95 K  and 110 K at $N_h$(IP)$\sim$0.15 and $\sim$0.16 (see Fig. \ref{fig:Knight_nlayer}(c)), respectively, any static AFM order has not been evidenced at their IPs even though these carrier densities are lower than $N_h=$0.169 at QCP for five-layered compounds. 
Remarkably, the extremely short spin-spin relaxation time was observed in the case of $N_h$(IP)$\sim$0.15 for underdoped Hg-1234, suggesting the closeness to the QCP of four-layered cuprates.
These results strongly suggest that the QCP moves to a region of lower carrier density for $n$-layered cuprates with $n \le 4$, as illustrated in Fig. \ref{fig:phase_nlayer}. 
Although the AFM superexchange interaction among spins at the nearest neighbor Cu sites in a CuO$_2$ plane is as large as $J_{ab}\sim$ 1300 K \cite{Tokura}, the effective interlayer coupling depends on the structural details and number of CuO$_2$ planes. Assuming an isolated two-dimensional (2D) system, any long-range AFM order is not expected at a finite temperature. Therefore, this result reminds again that the interlayer coupling is crucial for the onset of AFM order. In a five-layered system, three underdoped IPs may stabilize the long-range AFM order due to sufficient interlayer coupling. In this context, it is the weak interlayer magnetic coupling that suppresses the AFM order in LSCO \cite{Keimer} and YBCO \cite{Sanna} at such small carrier densities region. 

\subsection{Proposal for understanding of underdoped state }

As a result of the discussion above, we propose that the antiferromagnetically coupled spontaneous moment may persist up to $N_h$ $\sim$ 0.16 in the CuO$_2$ planes even in LSCO and YBCO as well, but may be hidden within the plane due to the strong 2D fluctuations brought about by the weak interlayer coupling. In fact, the application of a high magnetic field stabilizes the static AFM order in the vortex state in underdoped samples, but not in the optimally doped samples \cite{Lake,Miller}. Furthermore, the AFM order was also observed in the charge-stripe phases around $x\sim$1/8 of LSCO\cite{Tranquada}. As for the underdoped YBa$_2$Cu$_3$O$_{6.5}$ (YBCO6.5) with $T_{\rm c}$ = 60 K, the AFM order has been reported in a neutron-scattering experiment, which suggested that an AFM moment of 0.1$\mu_{\rm B}$ fluctuates in the nanosecond time scale \cite{Sidis}. In oxygen-ordered high-quality YBCO6.5, a quantum oscillation revealed the Fermi surface comprising a Fermi pocket \cite{dHvA}. This result may be understood by assuming that the Fermi surface is folded at the magnetic Brillouin zone ($\pi$,$\pi$) due to the presence of the AFM order under a very high field. The Fermi arc observed in the photoemission spectra of underdoped cuprates \cite{Norman} may also be explained by the Fermi pocket picture under the hidden short-range AFM order and by the collapse of a part of the Fermi surface caused by the very short life time of quasi-particles due to the disorder. Although the phase diagrams of LSCO and YBCO are widely believed thus far as typical phase diagram of cuprates, we claim that these underlying issues in their underdoped region may be affected by the strong 2D fluctuations produced by the weak interlayer coupling, in addition to the disorders caused by the chemical substitution for doping. 
This concept will lead us to a coherent understanding of underlying anomalies on underdoped cuprates, such as the AFM order induced by magnetic field \cite{Lake,Miller}, stripe order \cite{Tranquada} in LSCO at $x\sim1/8$, and the Fermi arc \cite{Norman} in a underdoped region.

\section{Conclusion}

The site-selective NMR studies on the five-layered cuprates have unraveled the genuine phase diagram of the homogeneously doped  CuO$_2$ plane: (1) the AFM order is robust up to $N_h \approx$ 0.17, (2) the uniform mixing of AFM order and SC takes place at least in $0.14 \le N_h \le 0.17$, (3) $T_{\rm c}$ has a peak close to the QCP at which the AFM order collapses, and (4) the tetracritical point for the AFMM/(AFMM+SC)/SC/PM phases may be present at $N_h \approx 0.15$ and $T \approx$ 75 K. These results suggest the intimate relationship between the high-$T_{\rm c}$ SC and AFM order, namely, the AFM superexchange interaction plays a vital role not only for the onset of the AFM order but also of SC. 
Although the phase diagrams of LSCO and YBCO  are widely believed thus far as typical ones, we claim that the underlying issues in underdoped LSCO, such as the stripe order, the magnetic-field induced AFM order, Fermi arc, etc, may be affected by the disorder and/or by strong 2D fluctuations due to the weak interlayer coupling, in addition to the strong correlation effect. The results presented here allow us to obtain an insight that the AFM superexhange interaction is the most promising glue for the Cooper pair in  cuprate superconductors.

\section*{Acknowledgement}

We would like to thank M. Mori, T. Tohyama, S. Maekawa, H. Kohno and M. Ogata for their discussions. This work was supported by a Grant-in-Aid for Creative Scientific Research (15GS0213) from the Ministry of Education, Culture, Sports, Science and Technology (MEXT) and the 21st Century COE Program (G18) of the Japan Society for the Promotion of Science (JSPS).



\begin{thebibliography}{99}
\bibitem{IPs} In this study, the IP in the central plane is also denoted as IP because it is similar to the other IPs due to the carrier density and local environment being same.
\end{thebibliography}
\end{document}